\documentclass[aip,reprint]{revtex4-2}
\usepackage{bm}
\usepackage{graphicx}
\usepackage{subfigure}
\usepackage{wrapfig}
\usepackage{epstopdf}
\usepackage{siunitx}
\usepackage{braket}
\usepackage{tabularx}
\usepackage{blindtext}
\usepackage{amsmath}
\usepackage{float}
\usepackage{tabularx}
\usepackage{multirow}
\usepackage{hyperref}
\usepackage{color}
\usepackage[section]{placeins}

\def\bem#1{\begin{mathletters}\label{#1}}
\def\eml{\end{mathletters}}

\def\4#1{{\boldsymbol{#1}}}
\def\8#1{{\widetilde{#1}}}

\makeatletter
\def\@email#1#2{%
 \endgroup
 \patchcmd{\titleblock@produce}
  {\frontmatter@RRAPformat}
  {\frontmatter@RRAPformat{\produce@RRAP{*#1\href{mailto:#2}{#2}}}\frontmatter@RRAPformat}
  {}{}
}%
\makeatother

\begin{document}

\title{Realization of robust quantum noise characterization in the presence of coherent errors}

\date{\today}

\author{P. Penshin $^*$}
\email{pavel.penshin@mail.huji.ac.il}
\affiliation{Dept. of Applied Physics, Rachel and Selim School of Engineering, Hebrew University, Jerusalem 91904, Israel}

\author{T. Amro $^*$}
\email{tamara.amro@mail.huji.ac.il}
\affiliation{Dept. of Applied Physics, Rachel and Selim School of Engineering, Hebrew University, Jerusalem 91904, Israel}

\author{T. Zabelotsky} 
\affiliation{Dept. of Applied Physics, Rachel and Selim School of Engineering, Hebrew University, Jerusalem 91904, Israel}

\author{A. Abramovich}
\affiliation{Dept. of Applied Physics, Rachel and Selim School of Engineering, Hebrew University, Jerusalem 91904, Israel}

\author{T. Pandit}
\affiliation{Fritz Haber Research Center for Molecular Dynamics, Hebrew University of Jerusalem, Jerusalem 9190401, Israel}

\author{K. I. O Ben 'Attar}
\affiliation{Dept. of Applied Physics, Rachel and Selim School of Engineering, Hebrew University, Jerusalem 91904, Israel}

\author{A. Hen}
\affiliation{Dept. of Applied Physics, Rachel and Selim School of Engineering, Hebrew University, Jerusalem 91904, Israel}

\author{R. Uzdin}
\email{raam@mail.huji.ac.il}
\affiliation{Fritz Haber Research Center for Molecular Dynamics, Hebrew University of Jerusalem, Jerusalem 9190401, Israel}

\author{N. Bar-Gill}
\affiliation{Dept. of Applied Physics, Rachel and Selim School of Engineering, Hebrew University, Jerusalem 91904, Israel}
\affiliation{The Racah Institute of Physics, The Hebrew University of Jerusalem, Jerusalem 91904, Israel}
\affiliation{The Center for Nanoscience and Nanotechnology, The Hebrew University of Jerusalem, Jerusalem 91904, Israel}

\begin{abstract}

Complex quantum systems and their various applications are susceptible to noise of coherent and incoherent nature. Characterization of noise and its sources is an open, key challenge in quantum technology applications, especially in terms of distinguishing between inherent incoherent noise and systematic coherent errors. In this paper, we study a scheme of repeated sequential measurements that enables the characterization of incoherent errors by reducing the effects of coherent errors. We demonstrate this approach using a coherently controlled Nitrogen Vacancy in diamond, coupled to both a natural nuclear spin bath (non-Markovian) and to experimentally controlled relaxation through an optical pumping process (nearly Markovian). Our results show mitigation of coherent errors both for Markovian and Non-Markovian incoherent noise profiles. We apply this scheme to the estimation of the dephasing time ($T_2^*$) due to incoherent noise. We observe an improved robustness against coherent errors in the estimation of dephasing time ($T_2^*$) compared to the standard (Ramsey) measurement.

\end{abstract}

\maketitle

\section{Introduction}
Quantum technologies, ranging from quantum information processing and quantum computing to quantum sensing, have developed tremendously over recent years, with the growing complexity of quantum systems leading to advanced capabilities \cite{devoret2013superconducting}\cite{awschalom2013quantum}.
The practical utilization of such complicated quantum systems is usually realized through analysis and control of simpler building blocks, namely two-level systems (TLSs), commonly known as qubits \cite{nielsen2010quantum}\cite{monroe2013scaling}\cite{degen2017quantum}.
Such TLSs can be implemented using various physical platforms, each characterized by its quantum properties, control techniques and other considerations \cite{chuang1998quantum}\cite{knill2000theory}.

Practical quantum TLSs, and even more so the complex quantum systems built upon them, are prone to environmental noise, posing a significant challenge for achieving high-fidelity quantum operations. The characterization of such environmental noise, which we refer to as incoherent errors, is a crucial aspect in enabling useful quantum technologies with any quantum platform. Moreover, understanding and characterizing such incoherent errors could suggest strategies to mitigate their effects and enhance the coherence and robustness of the quantum system. However, most techniques used to study incoherent errors (sometimes referred to as noise spectroscopy) rely on coherent control of the quantum system, which in itself is not perfect. We denote the imperfections in the unitary evolution resulting from the control as coherent errors.
Other methods based on random circuits, known as randomized benchmarking \cite{PhysRevLett.126.210504}and its variants \cite{RB_Coherent_v_incoherent} suffer from scalability issues\cite{PRXQuantum.3.020357} \cite{PhysRevLett.129.150502}\cite{Cross_2016}.

Due to the complexity of quantum evolution it is very difficult to distinguish between coherent and incoherent error as presently there are no tools that can accomplish this task in a scalable manner. An alternative that overcomes the scalability problem is restricted to the weak noise regime\cite{santos2023scalable}.

Here we experimentally explore a scheme aimed at mitigating the impact of coherent errors in the context of characterizing incoherent errors, as well as attempting to estimate purity loss in the low action regime. This approach is based on the theoretical development of  \cite{santos2023scalable}, and is demonstrated on a simple quantum TLS realized using a nitrogen-vacancy (NV) center in diamond.

In this paper, we investigate the sensitivity of the proposed scheme to deviations from Markovianity in both nearly Markovian and non-Markovian noise environment regimes (over which we have experimental control) and obtain insights into the validity regime of this approach.

The quantum system employed experimentally in this work is based on the NV center
in diamond\cite{Doherty2013}, which possesses appealing qualities such as room temperature operation, optical accessibility, and long coherence times, making it a highly attractive platform for various quantum applications 
 \cite{pham2011} \cite{mcguinness2011}\cite{bouchard2011}. 
The NV is prone to environmental noise mostly through coupling to magnetic noise from surrounding spin defects \cite{Chrostoski_2021}, as well as through thermal noise associated with phononic excitations in the diamond lattice \cite{Cambria_2021}.

We realize a scheme for characterizing incoherent errors while mitigating effects of coherent errors, based on the repeated application of the (imperfect) unitary control \cite{santos2023scalable}.We employ a linear sum of $ n $  state measurements, where each measurement is obtained by repeating the same unitary evolution $k$ times
\begin{equation}
\sigma_n =\sum_{k=0}^n a_k^{(n)} R_k. \\
\label{eqn:General Sn}
\end{equation}
In the expression above $R_k = \text{tr}[\rho_0 \rho_k]$ is used to quantify the fidelity between the initial state $\rho_0$ and the final state $\rho_k$ (when $\rho_0$ is a pure state), which is achieved after performing the unitary evolution k times. $R_0$ is the fidelity of the initial state with itself, for a pure state $R_0 = 1$. $ R_1$ is measured in the presence of coherent and incoherent noise for a specific duration, and similarly, $R_2$ is measured for twice that time, and so on. To minimize the system's sensitivity to coherent errors, one can take higher orders of $n$ in (\ref{eqn:General Sn}), leading to a more accurate characterization of the  incoherent noise within the system.\\ The weights in the sum $a_k^{(n)}$ correspond to the second row of the inverse Vandermonde matrix \cite{santos2023scalable} as follows:

\begin{equation}
a_k^{(n)}=\left(V_n^{-1}\right)_{2, k+1}
\end{equation}

In this work we focus on $n =2,3$, namely:
\begin{equation}
\sigma_2  =-\frac{3}{2} R_0+2 R_1-\frac{1}{2} R_2, \\
\label{sigma_2}
\end{equation}
\begin{equation}
\sigma_3  =-\frac{11}{6} R_0+3 R_1-\frac{3}{2} R_2+\frac{1}{3} R_3, \\
\label{sigma_3}
\end{equation}
It can be shown that taking higher orders of $R_k$ leads to a more accurate characterization of the incoherent noise within the system. Yet, the inclusion of higher order $R_k$'s entails an increase in the runtime of the protocol. This is a result of both the increased runtime with larger $k$ (due to the longer evolution time), as well as since the variance of $\sigma_n $ increases with higher orders of $n$ thus requiring more shots, which means longer experiments. 

The  dimensionless time Markovian evolution operator in Liouville space is (\ref{eqn:K}):
\begin{equation} 
\label{eqn:K}
K=e^{-i \mathcal{H}+\mathcal{L}} = e^x
\end{equation}
Where $\mathcal{H}$ generates the unitary part  of the evolution  and the dissipator  $\mathcal{L}$ generates the non-unitary part of the evolution. As derived in \cite{santos2023scalable}, the incoherent infidelity can be estimated through:  
\begin{equation}
\begin{aligned}
    \sigma_2 = \langle \mathcal{L} \rangle + O(x^3),
    \\
    \sigma_3 = \langle \mathcal{L} \rangle + O(x^4).
\end{aligned}
\end{equation}

\section{Methods}
We test the robustness of  $\sigma_2, \sigma_3$ to different levels of coherent errors (driving) and different levels of markovianity (of the incoherent noise) on a well-controlled qubit realization using an NV quantum spin in diamond \cite{Doherty2013}.

The NV center is a point defect in diamond that consists of a substitutional nitrogen atom and an adjacent vacant lattice site. The negatively charged NV\textsuperscript{-} is a spin $ S=1 $ system. The spin state of the NV center can be optically initialized and read out. Coherent control of the spin state is achieved by the application of a microwave (MW) drive. Separate spin sub-manifolds, $m_s = 0 \leftrightarrow m_s = 1 $  or $m_s = 0 \leftrightarrow m_s = -1 $, can be selectively addressed by exploiting the Zeeman splitting under an externally applied magnetic field. Therefore, the NV center can be treated as a two-level system in the context relevant to this work.

The NV center is exposed to noise generated by the surrounding environment. Specifically, our system consists of an ultrapure diamond sample with isotopically purified carbon content (Element Six: $<1$ ppb nitrogen and $99.99\%$ $^{12}$C), implanted with relatively deep, low concentration NVs (20 keV implantation energy, $10^{10}$ [cm $^{-2}$] dose). For this sample the main source of incoherent noise leading to dephasing of our NV quantum spin (determining its dephasing time $T_2^*$ \cite{childress2006}) is spin-carrying nuclei (\textsuperscript{13}C) that form a (relatively dilute) spin bath in the diamond lattice. This spin bath is a source of effectively non-Markovian noise due to its extended memory time, resulting from slow dynamics of such a nuclear spin system \cite{Haase_2018}.

\begin{figure}[tbh]
    \centering
    \includegraphics[width = 1 \linewidth]{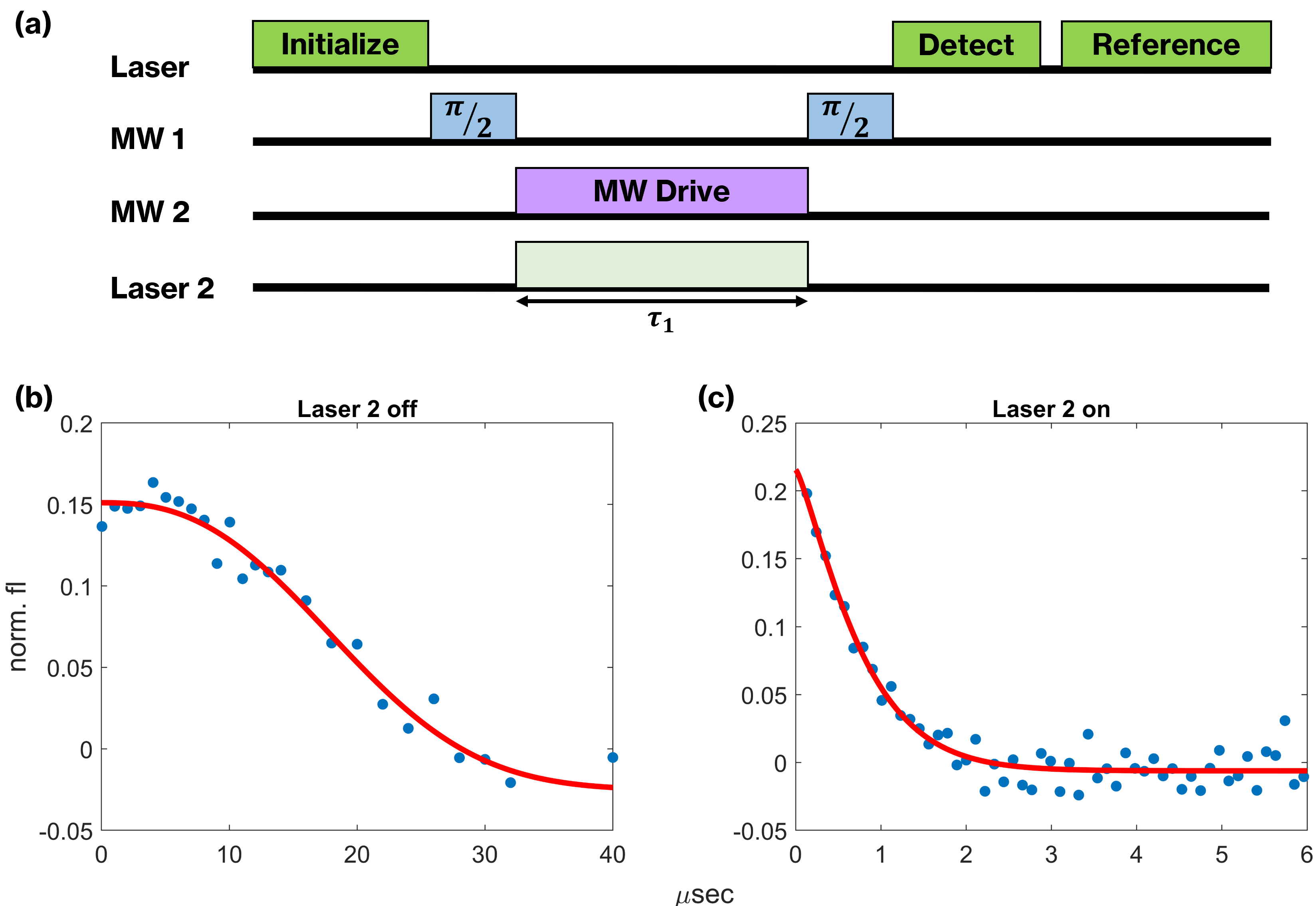}
    \caption{(a) Driven Ramsey sequence. Initialization, detection, and reference laser power are 309 $\mu W$, and Laser 2 power is $34.6 \mu W$. Laser 2 is used for the creation of nearly Markovian noise conditions. (b) Results for non-driven Ramsey experiment under Non-Markovian noise conditions (Laser 2 off). (c) Nearly Markovian noise conditions are reproduced when laser 2 is on. Application of the laser during the FID shortened the dephasing time $T_2^{*}$  from 22.1 $\mu s$ to 0.81 $\mu s$, as well as affected the power exponential behavior described in Eq.(\ref{eqn:De_sousa}). This power in (c) is $r\approx 1.23$ and in (b) $r\approx 2.47$.}
    \label{fig:Methods_Fig}
\end{figure}

The realization of the $\sigma_2$ and  $\sigma_3$ schemes was performed through standard dephasing measurements using Ramsey (FID - free induction decay) sequences \cite{childress2006}, which embody the canonical characterization of static (inhomogeneous) noise through the dephasing time $T_2^*$. This standard scheme is complemented here by an additional MW driving (MW2) and an optional low-power laser pumping pulse during the free induction decay [see Fig. \ref{fig:Methods_Fig}(a)].

The role of the MW2 driving is to induce a small coherent error in each cycle - rotations determined by the strength of MW2 drive. The fidelities $ R_1 $, \,$ R_2, \,R_3 $ and $ R_0 $ required for calculating $\sigma_2$ and $\sigma_3$ in Eq.(\ref{sigma_2},\ref{sigma_3}) can be extracted from the driven Ramsey sequence: applying the MW2 drive for $\tau_1 \rightarrow 0$ results in $R_0$ ; applying MW2 for fixed times of $\tau_1,\,2 \tau_1,\, 3 \tau_1$ generates $R_1$, $R_2$ and $R_3$ respectively. In this experimental realization, the phase between the MW1 and MW2 sources was locked and set to $0^\circ$, so as to ensure consistent control and coherent rotations, with the "coherent errors" introduced by MW2 producing rotations around the same axis (on the Bloch sphere) as the initial and final $\pi/2$ pulses produced by MW1. 

As mentioned above, the main noise source in our system originates from the surrounding nuclear spin bath of $^{13}$C isotopes, characterized by non-Markovian dynamics.
The low-power laser pumping was introduced in our experiments to tune the environmental noise toward a predominantly Markovian nature (short memory). This is achieved through a controlled optical pumping process, through which a nearly Markovian relaxation is realized \cite{Doherty2013}, the rate of which is determined by the intensity of this optical pumping pulse.

The degree of (non-) Markovianity of the noise can be characterized by the power parameter $r$ of the decaying stretched-exponential depicting the fidelity of the initial quantum state as a function of time in a Ramsey experiment \cite{sousa2009}, shown in Eq. \ref{eqn:De_sousa}.
\begin{equation}
    f(t) = \exp\left[- \left( t/T_2^* \right)^r \right].
    \label{eqn:De_sousa}
\end{equation}

In Eq. \ref{eqn:De_sousa} the stretched exponential decays with a power of $r\rightarrow 1 $ if the noise affecting the system is dominantly Markovian, while higher values of $r$ would imply that the system is non-Markovian (up to a power of 2 for long memory-time noise \cite{Bar-Gill2013,sousa2009}, see Fig. \ref{fig:Methods_Fig}(b)).

The introduction of the low-power laser pumping during the FID step is shown in Fig.\ref{fig:Methods_Fig}(c); this results in a nearly normal exponential decay curve, consistent with the Markovian noise regime. In contrast, when Laser 2 is not applied the non-Markovian environmental noise (nuclear spin bath) is dominant. Therefore, by applying Laser 2 we can control the Markovianity of the noise that affects our system. The added Markovian decay using the laser effectively shortens the $T_2^*$ dephasing time of the system as shown in Fig. \ref{fig:Methods_Fig}(b,c).

The experiments were performed using a home-built scanning confocal microscope equipped with a 532 nm laser source and an SPCM (single photon counting module) for collecting the fluorescence. Control over the NV spin state was performed using the application of a microwave signal generated by a signal generator, modulated by an IQ mixer, and delivered via a coplanar waveguide. Additional details of the experimental setup can be found in Appendix A.

The experimental data was complemented by complete simulations of the realistic quantum dynamics, including the time-dependent coherent control and the readout noise relevant to our system. The simulations included an effective model for the environmental noise, established through a stochastic time-dependent process based on the Ornstein–Uhlenbeck model \cite{OUprocess}, capable of capturing both Markovian and non-Markovian characteristics of the environment \cite{Gillespie}.

\section{Results}

\begin{figure}[tbh]
    \centering
    \includegraphics[width = 1 \linewidth]{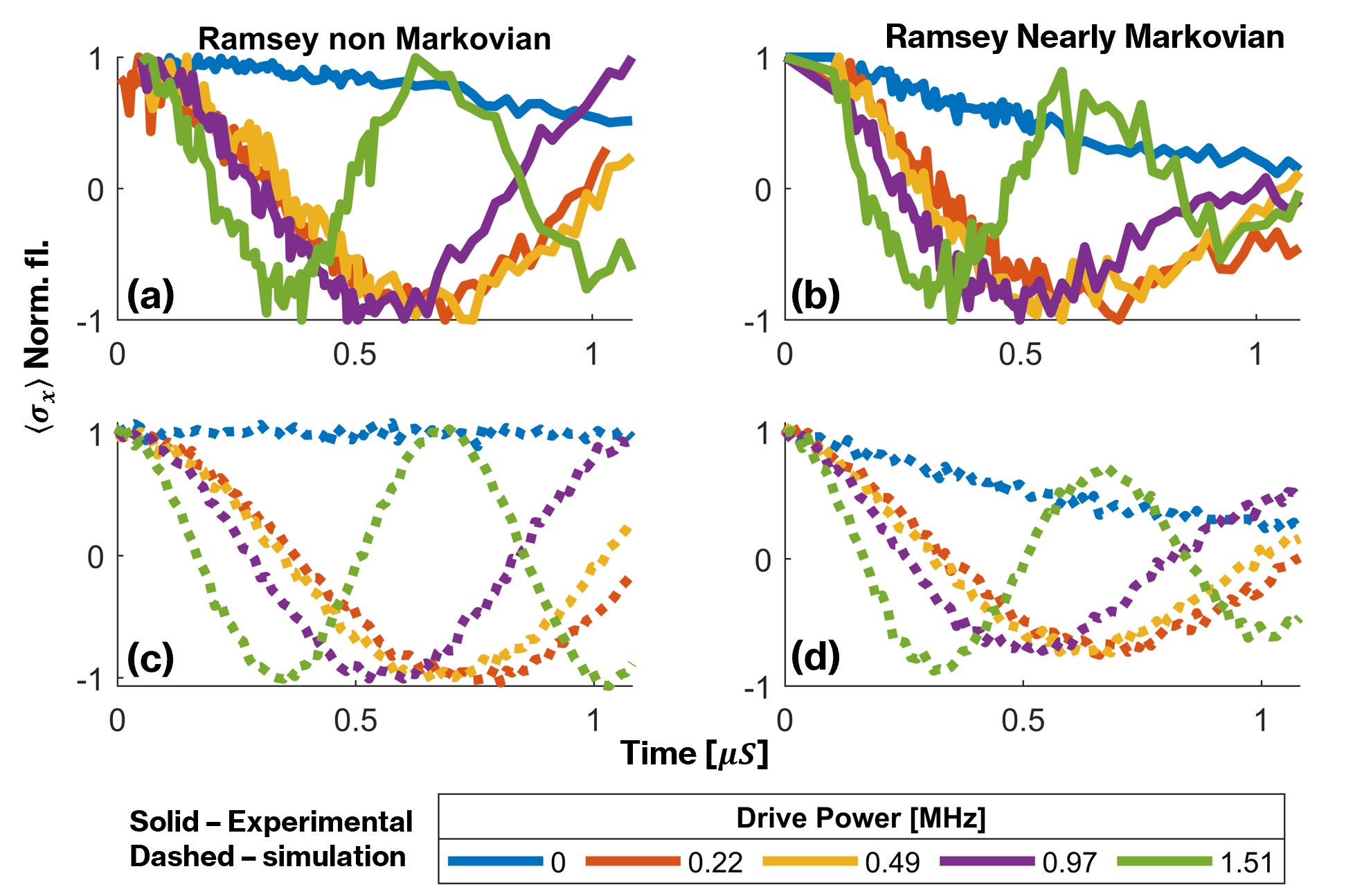}
    \caption{Results for the driven Ramsey experiments (upper row): (a) non-Markovian noise and (b) nearly Markovian noise. Corresponding simulations (bottom row): (c) non-Markovian, (d) nearly Markovian. The horizontal axis is the varying duration of the MW2 drive. Results for non-Markovian and nearly Markovian noise conditions are on the left and right columns respectively. The nearly Markovian regime was achieved by applying the same power of Laser 2 as in Fig. \ref{fig:Methods_Fig}(b).}
    \label{fig:Driven_Ramsey_Results}
\end{figure}

The sequence described in Fig.\ref{fig:Methods_Fig}(a) 
was performed on the same NV that yielded the $T_2^*$ results depicted in Fig. \ref{fig:Methods_Fig}(b,c). The experimental conditions for realizing nearly Markovian and non-Markovian noise regimes were as described above. The results are presented in Fig. (\ref{fig:Driven_Ramsey_Results}), and show an oscillating behavior of the measured fidelity $\left\langle \sigma_{x}\right\rangle$. These oscillations depend on the power of the applied MW2 drive during the free induction decay, which models coherent errors (similar to Rabi oscillations).

\begin{figure}[tbh]
    \centering
    \includegraphics[width = 1 \linewidth]{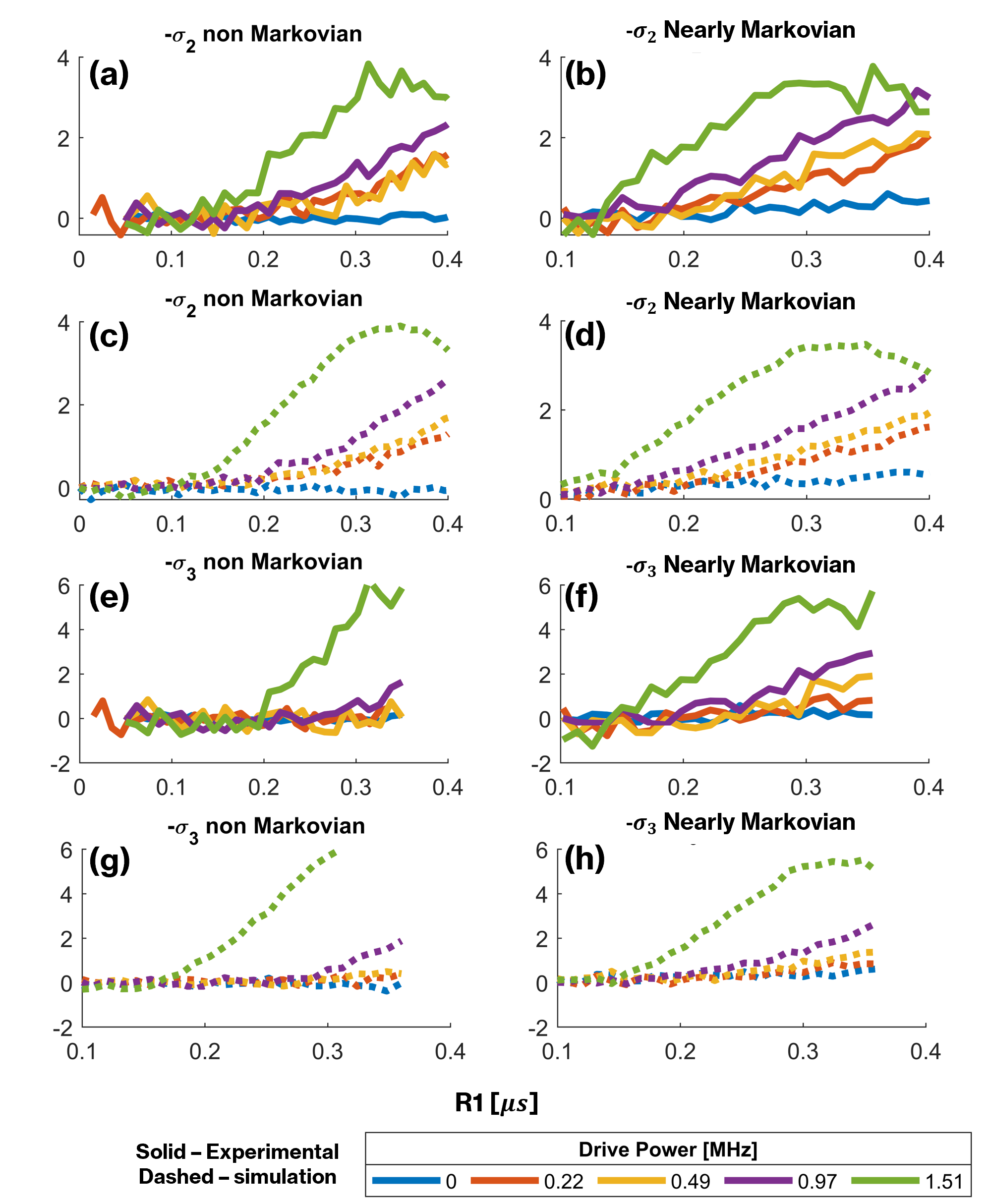}
    \caption{Extracted $\sigma_2$ and $\sigma_3$ values from the time dependent data/simulations presented in Fig. (\ref{fig:Driven_Ramsey_Results}). The horizontal axis is the value of the $R_1$ pulse time. Experimental results of $\sigma_2$ for non-Markovian and nearly Markovian noise are shown in (a) and (b), and the corresponding simulations are shown in (c) and (d). Similarly, the experimental results for $\sigma_3$ are depicted in (e),(f), while the corresponding simulations in (g),(h).
    The simulations agree well with the experimental data. Note that the lowest $R1$ value of the experimental data for the nearly Markovian part is limited by pulse delays between the Laser 2 and MW signals in the setup.}
    \label{fig:S2_and_S3_Results}
\end{figure}

Similarly to the results in Fig. \ref{fig:Methods_Fig}(b,c), the decay for the nearly Markovian regime is visibly faster due to the application of optical pumping (which is chosen to be dominant for these experiments).

The collected data from the driven Ramsey experiments were used to construct the $\sigma_2$ and $\sigma_3$ values using Eqs(\ref{sigma_2},\ref{sigma_3}). These $\sigma_2$ and $\sigma_3$ values are presented in Fig. (\ref{fig:S2_and_S3_Results}). 

As described in \cite{santos2023scalable}, the $\sigma_2$ and $\sigma_3$ schemes are expected to be less sensitive to coherent errors compared to regular Ramsey measurements. This increased robustness can be qualitatively identified from a comparison between Figs. (\ref{fig:Driven_Ramsey_Results}) and (\ref{fig:S2_and_S3_Results}).

\begin{figure}[tbh]
    \centering
    \includegraphics[width = 0.95 \linewidth]{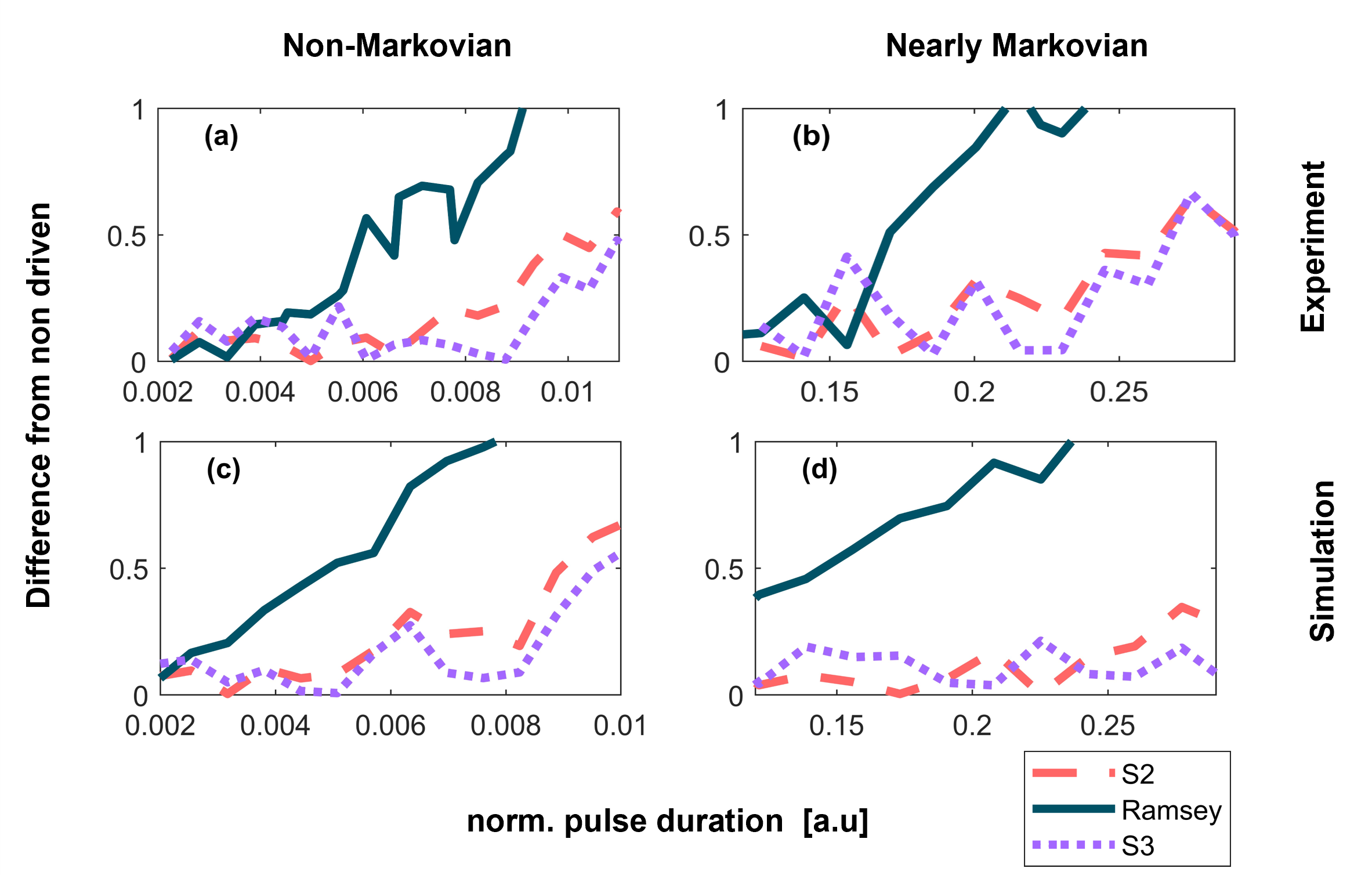}
    \caption{Accumulated error as a function of time, with respect to the non-driven (no coherent errors) case. Results for experiments (a, b) and simulations (c ,d), for non-Markovian (without optical pumping, a,c) and nearly Markovian (b,d) noise conditions. A single representative drive power (level of coherent error) of 2.09 MHz is presented. The pulse time axis for each noise regime was normalized by the corresponding value of $T_2^*$. }
    \label{fig:Sensitivity_Results}
\end{figure}

In order to quantitatively compare the robustness to coherent errors of the $\sigma_2$, $\sigma_3$ methods to standard dephasing (Ramsey) measurements, we examined the difference between the driven (with coherent errors) and non-driven cases for all schemes: Ramsey, $\sigma_2$ and $\sigma_3$. This difference is presented as a function of the normalized pulse length with respect to $T_2^*$, to overcome the effect of $T_2^*$ shortening between the nearly Markovian and non-Markovian regimes. This analysis is presented in Fig. (\ref{fig:Sensitivity_Results}).  
Our results indicate that the $\sigma_2$ $\sigma_3$ schemes are indeed consistently less sensitive to coherent errors compared to Ramsey measurements. The accumulated error with respect to the non-driven case in Ramsey increases more steeply compared to $\sigma_2, \sigma_3$, for both nearly Markovian and non-Markovian noise regimes.

Beyond the clear quantitative advantage depicted in Fig. \ref{fig:Sensitivity_Results}, it is important to analyze the usefulness of this robustness to coherent errors in the context of characterizing the fundamental incoherent errors. In our case the relevant quantifier is our ability to extract the value of $T_2^*$ from the dynamics in the presence of increasing coherent errors. We focus on the weak coherent error regime, realized either by working at short times or with low values of drive powers. 

In the following we rely on the correspondence between our experimental results and the simulations, as presented in Figs. \ref{fig:Driven_Ramsey_Results} , \ref{fig:S2_and_S3_Results}. Due to technical limitations associated with short pulses and low drive powers, we employ simulations to quantify $T_2^*$ at various low drive powers. The simulations take into account the same conditions as the experimental data in Fig.(\ref{fig:Methods_Fig}). The dephasing time $T_2^*$ was extracted by fitting the dynamics of Ramsey, $\sigma_2$ and $\sigma_3$ using the corresponding equations (\ref{eqn:Ramsey_fit_model}) , (\ref{eqn:S2_fit_model}), (\ref{eqn:S3_fit_model}):

\begin{equation}
    f_{Rams.}(t)=\exp\left[-\left(t/T\right)^{r}\right]
    \label{eqn:Ramsey_fit_model}
\end{equation}
    
\begin{equation}
    f_{\sigma_{2}}\left(t\right)=\frac{3}{2}-2\exp\left[-\left(t/T\right)^{r}\right]+\frac{1}{2}\exp\left[-\left(2t/T\right)^{r}\right]
    \label{eqn:S2_fit_model}
\end{equation}
\begin{multline}
    f_{\sigma_{3}}\left(t\right)=\frac{5}{3}-\frac{5}{2}\exp\left[-\left(t/T\right)^{r}\right] +\exp\left[-\left(2t/T\right)^{r}\right] - \\ \frac{1}{6}\exp\left[-\left(3t/T\right)^{r}\right]
    \label{eqn:S3_fit_model}
\end{multline}

\begin{figure}[tbh]
    \centering
    \includegraphics[width = 0.8\linewidth]{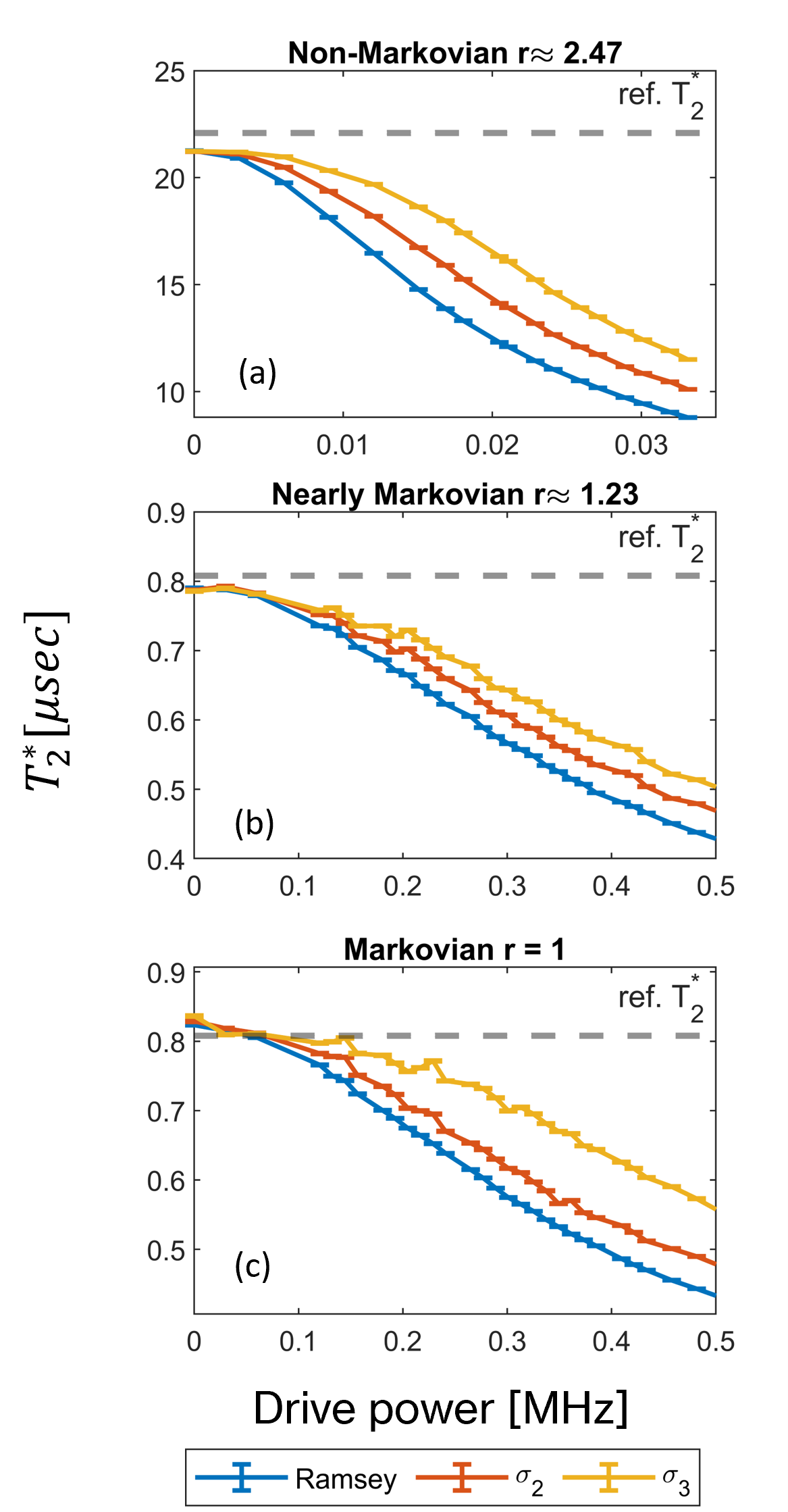}
    \caption{Extracted $T_2^*$ from low drive Ramsey, $\sigma_2$ and $\sigma_3$ dynamics (simulations). The reference $T_2^*$ values are the same values from Fig.(\ref{fig:Methods_Fig}).}
    \label{fig:T2_fitting}
\end{figure}

The simulations for each drive power were repeated multiple times and fitted to the corresponding models. The mean value and the standard error of the mean were used to determine the $T_2^*$ value for each drive power as shown in Fig. (\ref{fig:T2_fitting}). We note that the range of powers differs between the nearly Markovian ($r \approx 1.23$) and non-Markovian ($r\approx2.47$) experimental cases, as a result of the shortening of $T_2^*$ (for Markovian vs. non-Markovian noise), and the intent to keep the dynamics in the low action range. We also added a simulation of fully Markovian noise, with r=1. 

The results in Fig.(\ref{fig:T2_fitting}) indicate that the presence of increasing coherent errors causes an increasing deviation in the estimation of the $T_2^*$ for all noise regimes. Importantly, this effect is mitigated by $\sigma_2$ and more so by $\sigma_3$, corroborating the theoretical finding in \cite{santos2023scalable} for the fully Markovian case. The mitigation also happens for the non-perfect Markovianity that was experimentally achieved, as shown in Fig. \ref{fig:T2_fitting}(b) for $ r = 1.23$. The mitigation by $\sigma_2$ and $\sigma_3$ becomes even more pronounced when the system is closer to a perfect Markovian case, as shown in Fig. \ref{fig:T2_fitting}(c) for the case of $r = 1$.

We note that beyond its ability to mitigate coherent error effects in the estimation of incoherent noise, the $\sigma_n$ scheme is furthermore a good estimator of the purity loss of the system in the weak action regime \cite{Pandit2022boundsrecurrence}.

\begin{figure}[tbh]
    \centering
    \includegraphics[width = 1.1 \linewidth]{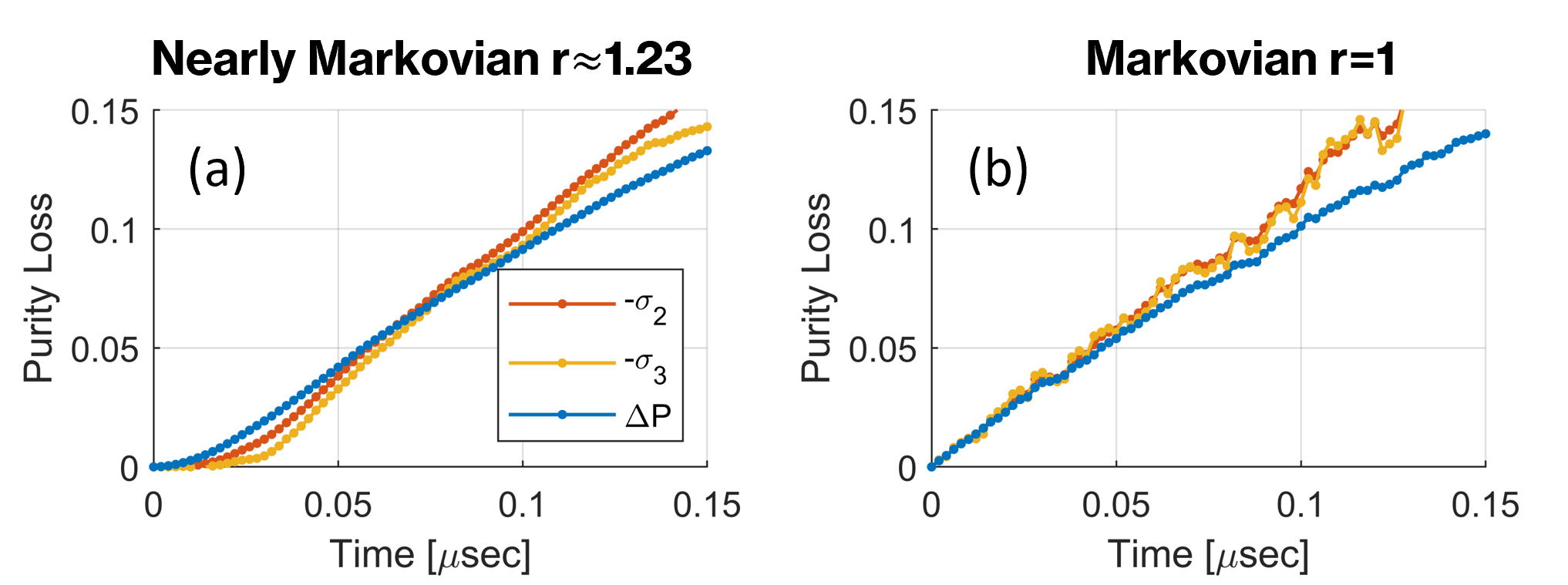}
    \caption{Comparison of simulation results between purity loss $\Delta P$  by using tomography and $- \sigma_2 , - \sigma_3$ at 0.3 MHz drive. (a) Simulation of nearly-Markovian conditions as were realized experimentally. (b) Simulation of fully Markovian conditions, depicting improved reproduction of purity loss by the $-\sigma_3 ; -\sigma_2$ schemes. }
    \label{fig:Purity_loss}
\end{figure}

In Fig. \ref{fig:Purity_loss} we plot the simulated purity loss vs. time, in the presence of Markovian noise and in the weak action regime. The purity loss can be estimated by $\sigma_n$  as follows (see appendix c): 
\begin{equation}
\Delta P = 1 - Tr(\rho^2) \cong \sigma_n,   
\end{equation}
where $ \cong $ represents equality in the weak action regime. We extract this measure as a function of time, using the simulated data and limiting to the weak action regime at $< 5\%$ purity loss. For simplicity and visibility, this simulation does not include measurement noise.
These results indicate that $\sigma_n$ measurements in the low action regime follow the purity loss trend, with accuracy improving as the evaluated system approaches fully Markovian dynamics with $r=1$ (instead of the experimentally realized nearly-Markovian noise with $r \approx 1.23$). The potential advantage of the presented method is the ability of fast characterization of the purity loss in Markovian systems, instead of performing longer experiments based on state tomography.

\section{Conclusions}
In this study, our objective was to assess the sensitivity of a recent method designed for measuring infidelity to the level of markovianity and to the level of coherent error. Our findings reveal significant sensitivity to deviations from markovianity. Nevertheless, even in scenarios where the method might not capture ( the non-Markovian case ) the desired $T^*_2$ information, it still demonstrates the suppression of coherent errors compared to the simple FID experiment. A similar trend was also observed when simulating purity loss in the nearly-Markovian case. 

We conclude that while this method can be confidently applied in systems with well-established markovianity, extra caution is necessary in other systems, such as NV centers, where Non-markovian effects are not negligible. 

With this in mind, the appealing scalability of the method makes it a potent candidate for quantifying incoherent errors in quantum computers and metrology applications, where coherent noise poses a substantial obstacle to achieving high-fidelity quantum operations. 

\section*{Acknowledgments}
\section*{Author declarations}
\subsection*{Conflict of Interest}
The authors have no conflicts to disclose.
\subsection*{Author Contributions}
\section*{Data Availability}
The data that support the findings of this study are available
from the corresponding author upon reasonable request.
\section*{Appendix A - experimental setup}

The experimental setup consisted of a confocal microscope setup using a Nikon Plan Apochromat objective, magnification x100 with NA of 1.45. The delivery of MW pulses was done using SRS SG396 signal generator, amplified using a Mini-Circuits amplifier ZHL-5W-63S+. The signal was delivered to the diamond sample using a coplanar waveguide. The NV fluorescence is filtered with a 650nm long-pass filter and collected using Excelitas SPCM. The MW source drive power is measured in units of the TLS Rabi frequency while making sure that all experiments are performed without detuning. 
\begin{figure}[tbh]
    \centering
    \includegraphics[width = 0.7\linewidth]{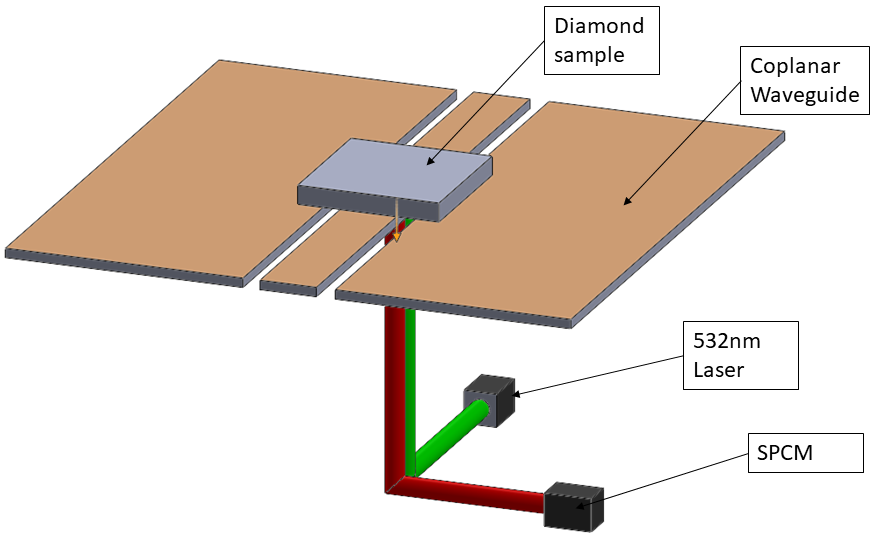}
    \caption{Schematic of the experimental setup, consisting of a home built confocal fluorescence microscope, in which green excitation and red fluorescence are combined. The diamond sample is mounted on a microwave stripline antenna, which delivers the controlled driving.}
    \label{fig:system_Diagram}
\end{figure}

\section*{Appendix B - Simulation }
The simulation that was performed is an implementation of the Ornstein–Uhlenbeck process to simulate quantum bath magnetic noise proportional to  $\sigma_z$. The magnetic noise coefficients were calibrated to reflect the experimental conditions.

The state was initialized to $\left\langle \sigma_x \right\rangle = 1 $ and rotated around $\sigma_y$ during the "FID" phase to simulate the MW drive applied in the experiment. In most of the displayed results, we applied a measurement noise to the final signal to simulate the experimental shot noise. The simulated measurement noise power was calibrated to reflect the experimental noise power. State tomography could easily be done by calculating the trace of the state density matrix with the relevant spin operators. 

\begin{figure}[tbh]
    \centering
    \includegraphics[width = 0.7\linewidth]{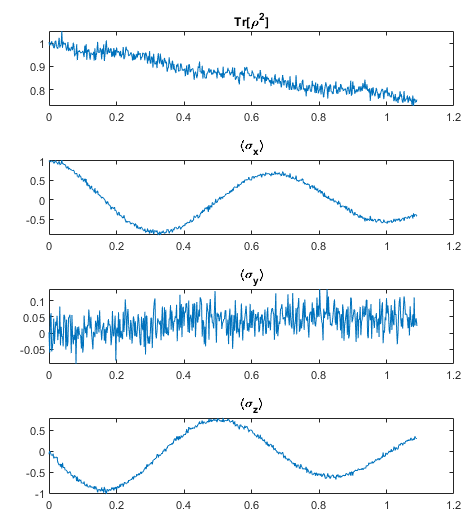}
    \caption{State tomography for Markovian conditions with MW driving.}
    \label{fig:system_Diagram}
\end{figure}

\section*{Appendix C -  derivation of purity loss relation to $\sigma_n$ and the case of perfect markovianity}

In a basic FID experiment, with no drive, after initialization into a $\sigma_{x}$, the eigenstate of the Bloch vector is 

\begin{equation}\tag{B.1}
\vec{a}=\left(\left\langle \sigma_{x}\right\rangle \left(t\right)\cdot\hat{x}+0\cdot\hat{y}+0\cdot\hat{z}\right) 
\end{equation}
 The $\sigma_{x}$ projection dynamics is described by simple exponential decay 
 \begin{equation*}\tag{B.2}
\left\langle \sigma_{x}\right\rangle \left(t\right)=\exp\left(\frac{t}{\tau}\right)    
 \end{equation*}. The purity for short times will be as follows: 
\begin{equation}\tag{B.3}
    Tr\left(\rho^{2}\right)=\frac{1}{2}\left(1+\left|\vec{a}\right|^{2}\right)=\frac{1}{2}\left(1+\exp\left(-\frac{2t}{\tau}\right)\right)\approx1-\frac{t}{\tau} 
\end{equation}
and purity loss will be $\Delta P=1-Tr\left(\rho^{2}\right)=t/\tau$.

Now let us look at $\sigma_{2}$ expansion at short times, this is valid for all $\sigma_{n}$ 
\begin{equation}\tag{B.4}
    \sigma_{2}=-\frac{3}{2}R_{0}+2R_{1}-\frac{1}{2}R_{2} 
\end{equation}
where $R_{0}=1\,\,\,R_{1}=\exp\left(-\frac{t}{\tau}\right)\,\,\,R_{2}=\exp\left(-\frac{2t}{\tau}\right)$
\begin{equation*}
\sigma_{2}=-\frac{3}{2}+2\exp\left(-\frac{t}{\tau}\right)-\frac{1}{2}\exp\left(\frac{2t}{\tau}\right)\approx
\end{equation*}
\begin{equation*}\tag{B.5}
    -\frac{3}{2}+\left(2-\frac{2t}{\tau}\right)+\left(-\frac{1}{2}+\frac{t}{\tau}\right)=-\frac{t}{\tau}
\end{equation*}

Thus we get $\Delta P = -\sigma_2 $ ,the same for $\sigma_3 $ expansion, the purity loss at short times is $\Delta P = -\sigma_3 $ as shown in Fig \ref{fig:Purity_loss}
\bibliography{references.bib}

\end{document}